\documentclass[lettrpaper,11pt]{article}
\usepackage[spanish]{babel}
\usepackage{latexsym}
\usepackage[dvips]{graphicx}
\usepackage[centertags]{amsmath}
\usepackage{amsfonts}
\usepackage{amssymb}
\usepackage{amsthm}
\usepackage[ansinew]{inputenc}
\usepackage{fancyhdr}
\usepackage{rotating}
\begin{document}

\begin{center}
\Large{\textbf{Exact Computation of Image Disruption Under Reflection on a Smooth Surface and Ronchigrams}}
\end{center}

\begin{center}
Edwin Rom\'an-Hern\'andez$^{1}$, Gilberto Silva-Ortigoza$^{2}$
\end{center}

\begin{center}
Facultad de Ciencias F\'isico
Matem\'aticas de la Universidad Aut\'onoma de Puebla, Apartado
Postal 1152, 72001, Puebla, Pue., M\'exico.
\end{center}

\begin{center}
rohe${\_00}$@hotmail.com$^{1}$, gsilva@fcfm.buap.mx$^{2}$
\end{center}

{\footnotesize{In this work we use geometrical optics and the \textit{caustic-touching theorem} to study, in an exact way, the change in the topology of the image of an object obtained by reflections on an arbitrary smooth surface. Since the procedure that we use to compute the image is exactly the same as that used to simulate the ideal patterns, referred to as ronchigrams, in the Ronchi test used to test mirrors, we remark that the closed loop fringes commonly observed in the ronchigrams when the grating, referred to as a Ronchi ruling, is located at the caustic place are due to a disruption of fringes, or, more correctly, as disruption of shadows corresponding to the ruling bands. To illustrate our results, we assume that the reflecting surface is a spherical mirror and we consider two kinds of objects: circles and line segments.}}

\section{Introduction}

By using the paraxial approximation of geometrical optics, Berry in an extraordinary work\,\cite{Berry}, among other things, studied the image of an arbitrary one-dimensional object obtained by reflection on an arbitrary smooth surface (of rippled water). He found that under certain conditions the object and its image do not have the same topology. To explain this beautiful phenomenon he introduced the so-called {\it caustic touching theorem}, which states that changes of image topology occur when the object touches the caustic associated with the family of imaginary light rays {\it emitted} by the observing eye.

\begin{figure}[h]
\begin{center}
\includegraphics[height=6cm]{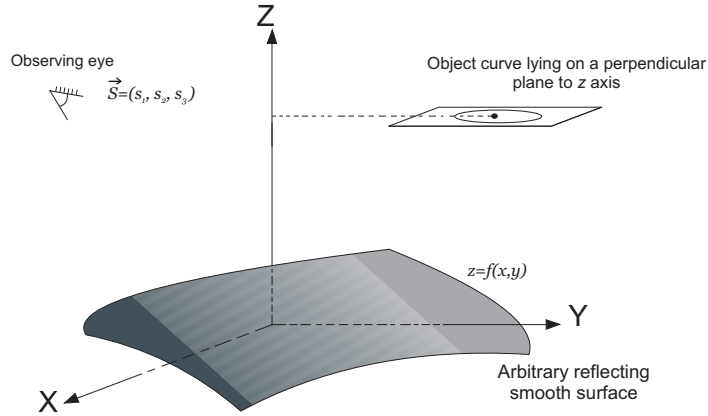}
\caption{\footnotesize Schematic drawing of the object curve, which we assume is lying on a plane perpendicular to the $z$ axis, the two-dimensional reflecting smooth surface locally given by $z = f(x, y)$ and the position of the observing eye given by $\vec{S} = (s_1, s_2, s_3)$.}\label{f:1}
\end{center}
\end{figure}

\begin{figure}[h]
\begin{center}
\includegraphics[height=6cm]{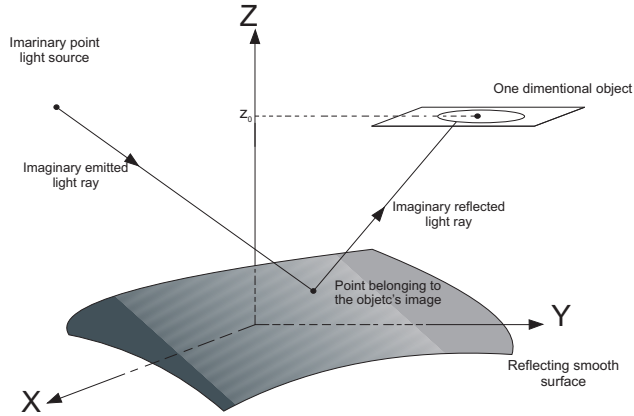}
\caption{\footnotesize The new geometrical arrange to compute the image of a one-dimensional object under reflection on the arbitrary smooth surface $z = f(x, y)$. The observing eye has been replaced by an imaginary point light source. In this diagram we show an imaginary emitted light ray and the corresponding reflected light ray which, we assume  arrives at a point of the one-dimensional object. Therefore, the point of the surface where the imaginary light ray is reflected belongs to the image of the one-dimensional object.}\label{f:2}
\end{center}
\end{figure}

It is worthwhile describing, with little convenient changes, the procedure followed by Berry to obtain the image of an arbitrary object curve by reflection on an arbitrary smooth surface. Without loss of the generality, we assume that: the object curve is lying on a plane perpendicular to the $z$ axis, the two-dimensional reflecting smooth surface is locally given by $z = f(x, y)$ and the position of the observing eye by $\vec{S} = (s_1, s_2, s_3)$, (see figure\,\ref{f:1}). At first instance (sight), one could think that to obtain the image of the one-dimensional object one has to take into account the family of light rays associated with each of its points. That is, a family of light rays characterized by three parameters (one of these parameters provides the position of an arbitrary point on the one-dimensional object and the other two give the direction of the light ray emitted from that point). However, the only important light rays that give contribution to the image formation are those that reach the observing eye and these, in accordance with the reciprocity principle\,\cite{Landau}, can be regarded as belonging to the single family emitted by the eye. In other words, the original problem of image formation is equivalent to a new problem where the observing eye is replaced by an imaginary point light source. From this new point of view, a point on the reflecting surface belongs to the image associated with the one-dimensional object if it can be associated, via a reflected light ray emitted by the imaginary point source, with a point of the one-dimensional object, as it is graphically described in figure\,\ref{f:2}. Since in general the curvature of the reflecting surface is not equal to zero, then the reflected light rays emitted by the imaginary point light source will focus at a region in the space. This region is the caustic associated with the imaginary reflected light rays. From a mathematical point of view, the evolution of the reflected light rays is described by a map between two subsets of $R^3$ or equivalently by a \textit{one-parameter family} of maps between two subsets of $R^2$ (see section\,2). The caustic associated with the reflected light rays is determined by looking for the points, on the reflecting surface, where these maps are not locally one to one. Therefore, if the one-dimensional object is placed outside the caustic region there will be a one to one correspondence between its points and the points of its associated image. That is, the observing eye will see only one image. In this case one says that the object and its image have the same topology. Remember that, roughly speaking, two curves  are said to be topologically equivalent if one can be transformed into the same shape as the other without connecting or disconnecting any points. However, if the object is located at the caustic region the maps are not one to one and its position with respect to the caustic will be \textit{crucial} to describe correctly its image. The caustic-touching theorem establishes that new image loops appear when the object curve touches the caustic associated with the reflected light rays emitted by the imaginary point light source. Such disruption may be elliptic, loop born from an isolated point, or hyperbolic, loop pinched off from an already existing one. In these cases, the object and its image do not have the same topology, and the observer may see several images corresponding to a single object. As remarked by Berry it is what happens in nature's optics, whose elements may be the reflecting surface of rippled water. The first aim of the present work is to obtain, within the geometrical optics approximation, an exact set of equations to study the change of  topology of a one-dimensional object obtained by reflection on an arbitrary smooth surface.

From the research developed by several authors\cite{Berry, Greenler, Fraser, Tape, Narayan, Hogan, Petters} it is quite clear that the multiple image formation process in the natural optical systems, whose components may be the reflecting surface of rippled water, refractive-index gradients in the atmosphere, or the gravitational field associated with a matter distribution, can be explained within the geometrical optics limit by using the caustic-touching theorem. On the other hand, the conventional optical systems such as microscopes and telescopes, whose basic components are mirrors and lenses, are designed to produce a single image, of course, as perfect as it is possible. So, at first glance, one could infer that non-multiple images, associated with a single object, can be seen by using these optical devices. However, in general, it is not true. The second and main contribution of this work is to remark that even with a perfect mirror it is possible, under certain conditions determined by the caustic-touching theorem, to observe image disruption. In particular, we remark that the caustic-touching theorem allow us to describe the pattern,  referred to as the ronchigram in the well known Ronchi test, when the grating is placed at the caustic region.

The Ronchi test\,\cite{Ronchi} developed by Ronchi in the 1920's is one of the most simple and powerful methods to extract information about the aberrations of an optical system. For these reasons it has been the subject of innumerable publications, from both the physical and the geometrical points of view (see Ref. \cite{Cornejo} and the references cited therein). From the physical point of view the fringes are interpreted as due to interference between several wavefronts produced by the ruling acting as a diffraction grating, while from the geometrical point of view the fringes are interpreted as shadows of the ruling slits. When the frequency of the ruling is not very high, the two points of view predict the same result. Since by analyzing the fringes of the real and ideal ronchigrams, one can determine the type and, in principle, the magnitude of the aberrations present at the exit pupil of the system, then it is fundamental to know the properties of the ronchigrams.

\begin{figure}[h]
\begin{center}
\includegraphics[height=6cm]{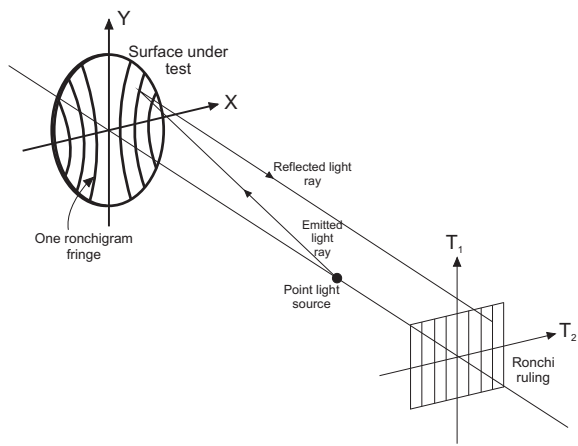}
\caption{\footnotesize Schematic drawing of the Ronchi test arrangement. In this diagram we have the surface under test, locally given by $z = f(x, y)$, a real point light source located on the optical system and a Ronchi ruling. The pattern observed through the grating on the surface of the mirror is referred to as the real ronchigram.}\label{RONCHITEST}
\end{center}
\end{figure}

In this work we consider the geometrical point of view of the Ronchi test. Under this assumption, the essential features of the Ronchi test for a concave mirror when the point source is located on the optical axis may be described by reference to figure\,\ref{RONCHITEST}. The light rays emitted by the point light source are reflected by the mirror under test and they focus to a region in the space. As previously mentioned, this region is the caustic associated with the reflected light rays. The grating, which is referred to as the Ronchi ruling, is located at different positions on the optical axis. The pattern observed through the grating on the surface of the mirror is referred to as the real ronchigram. From the geometrical point of view, the fringes of the real ronchigram are interpreted as shadows of the ruling bands. By comparing these real fringes with the ideal ones obtained by simulation one can deduce the defects of the mirror under test\cite{Malacara}. Therefore, one of the main  corner stones of the Ronchi method is the ideal ronchigram which is obtained by simulation. Thus, Sherwood\,\cite{Sherwood} has calculated the Ronchi pattern of a paraboloidal mirror when tested near the center of curvature; Malacara\,\cite{Malacara} developed an algorithm to predict the geometrical ronchigram of any spherical or aspherical mirror when tested at any point along the optical axis. These results have been generalized by Cordero, Cornejo and Cardona\,\cite{CorderoI} to an arbitrary mirror when the point source is located at any position, in particular their equations can be used to simulate ronchigrams for the cases of centered and off-axis conic sections with the point light source at any location. Cordero, D\'{\i}az, and Cabrera\,\cite{CorderoII} have presented a simple algorithm that allows the simulation of ronchigrams for any optical system in which it is possible to make an exact ray tracing. In accordance with the results obtained by these authors it is clear that the form and structure of the ideal ronchigram associated with a given perfect mirror depend on the location of both the point source and the Ronchi ruling. Furthermore, it is well known that when the Ronchi ruling is placed at the caustic region it is common to see closed loop fringes in the ronchigram. However, to our knowledge, an explanation of the closed loop fringes in the ronchigram using geometrical optics has not been presented. Actually, in practice for an easy interpretation of the ronchigram it is common to avoid any closed loop fringe, which is possible only when the grating is located outside  the caustic region associated with the reflected light rays\,\cite{Cornejo}. In this work we remark that the explanation is given by the caustic-touching theorem, and therefore, the relative position of the grating with respect to the caustic associated with the reflected light rays emitted by the point light source is \textit{crucial} to describe correctly the structure of the associated ronchigram. Our justification for these assertions is as follows. From figures \ref{f:2} and \ref{RONCHITEST}, it is clearly obvious that the procedure followed by Berry to compute the image of an arbitrary object under reflection on an arbitrary smooth surface when the observer is located at an arbitrary position in the space is exactly that used by several authors to simulate the ronchigram associated with an arbitrary smooth reflector by using a point light source located at different positions in the space. Remember that in Berry's procedure the observer is replaced by an imaginary point light source, if in addition, we replace the one-dimensional object by a grating, in particular a Ronchi ruling, then we get the desired configuration to simulate the corresponding ronchigram. Even though the simulation of the ronchigrams has been implemented for many years, to our knowledge, nobody has remarked on the fundamental role played by the caustic associated with the reflected light rays to describe the structure of the ronchigrams when the Ronchi ruling is located at the caustic region. Actually, as we will show in section 2, the equations that we obtain to compute the image associated with an arbitrary object obtained by reflection on an arbitrary smooth surface, are exactly those obtained by Cordero, Cornejo and Cardona\,\cite{CorderoI} in the context of the Ronchi and Hartmann tests. However, these authors did not study the properties of the ronchigrams when the Ronchi ruling is located at the caustic place.

An important observation is that in the Ronchi test the point light source is not an imaginary one, as in Berry's procedure, but it is a real one and the pattern, the image (in Berry's problem), is observed through the grating on the surface of the mirror. For practical purposes it is more convenient to plot the simulated ronchigram in a plane close to the reflecting surface. Therefore, in this work we plot the corresponding images (in Berry's problem) or patterns (in Ronchi's test) in the plane $z = 0$.

The aim of the present work is twofold: first, within the geometrical optics limit, we obtain an exact set of equations to study the changes of image topology under reflection on an arbitrary smooth surface and, second, we apply our results to two particular cases to show that even with a perfect spherical mirror one can observe the change of topology, in particular we remark that the structure of the ronchigrams when the Ronchi ruling is located at the caustic region can be described by using the caustic-touching theorem.

The organization of the present work is as follows: In section\,2, we use geometrical optics to obtain an exact set of equations to compute the image of a one-dimensional object produced by reflection on an arbitrary smooth surface. We find that this set of equations is exactly that previously obtained in the context of the Ronchi and Hartmann tests in\,\cite{CorderoI}. In section\,3, we compute the caustic surface associated with the light rays reflected by an arbitrary smooth surface when a point light source is located at an arbitrary position in the space. In section\,4, we collect the two sets of equations to study in an exact way the changes of image topology. Finally, in section\,5, we present some examples to illustrate the change of image topology when the reflecting surface is a spherical mirror and the objects are circles or line segments. In particular, we describe the structure of the ronchigram associated with a spherical mirror when the point light source is located on the optical axis and the Ronchi ruling is placed at the caustic region. \\

\section{ Computation of the image of a one-dimensional object obtained by reflection on an arbitrary smooth surface}

In this section, following Berry's procedure, we obtain the exact set of equations to compute the image of an arbitrary one-dimensional object obtained under reflection on an arbitrary smooth surface. Remember that the original problem is changed by a new one, where the observer is replaced by an \textit{imaginary} point light source as previously explained and shown in figure\,\ref{f:2}. Since this new problem is exactly that of simulating ronchigrams in the Ronchi test and there the point light source is a real one, from now on, we will use the term point light source to refer to both the real and imaginary point light sources, understanding that in the Ronchi test it is real and in the Berry's procedure  is an imaginary one. Thus assuming that in ${\cal R}^3$ we have a smooth arbitrary surface and a point light  source with arbitrary position we first obtain a parametric representation of the maps that describe the evolution of the reflected light rays by the arbitrary smooth surface, and then we obtain the set of equations that allow us to compute the image of a one-dimensional object obtained by reflection.

\begin{figure}[h]
\begin{center}
\includegraphics[height=6cm]{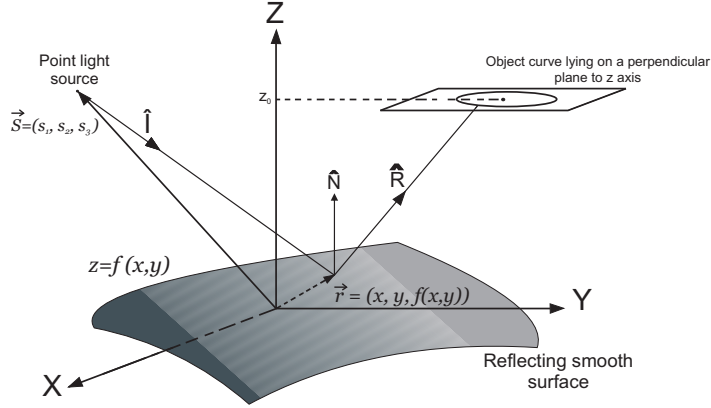}
\caption{\footnotesize Schematic drawing of the optical system and the vectors used to compute the image of a one-dimensional object obtained by reflection on an arbitrary smooth surface locally given by $z = f(x, y)$. In this diagram  $S = (s_1, s_2, s_3)$ denotes the position of the point light source,  $\hat{I}$ the direction of an emitted light ray, $\vec{r} = (x, y, f(x, y))$ the point on the smooth surface where the emitted light ray is reflected in the direction $\hat{R}$ and $\hat{N}$ is the normal vector to the smooth surface at the point of reflection.}\label{f:4}
\end{center}
\end{figure}

If the point light source is located at ${\vec S} = $ ($s_1, s_2, s_3$) and the smooth arbitrary surface is locally given by $z = f(x, y)$, see figure \ref{f:4}, then the light ray reflected at
the point ${\vec r}= $ ($x, y, f$($x, y$)) is described by
\begin{equation}
{\vec T} =  {\vec r} + l {\hat R},
\end{equation}
where $l$ is the distance along the reflected light ray, ${\hat R}$ is given by
\begin{equation}
{\hat R} = {\hat I}-2 ({\hat I} \cdot  {\hat N}){\hat N},
\end{equation}
${\hat N}$ is the unit normal vector to the reflecting surface  $z = f(x, y)$, and $\hat{I}$ gives the direction of the diverging ray from the point light source. From figure\,\ref{f:4} we have
\begin{equation}
{\hat I}   =  {\frac {\vec I}{|\vec I|}} =  \frac {(x - s_1, y
- s_2, f - s_3)} {\sqrt{(x - s_1)^2 + (y - s_2)^2 +
(f - s_3)^2}}.\\
\end{equation}

In order to obtain a vector field perpendicular to the reflecting surface, we define the function  $G(x,y,z) = z - f(x,y)$. Observe that one level surface of this function is the reflecting surface $z = f(x, y)$. Therefore, a vector field perpendicular to the reflecting surface is given by
\begin{equation}
{\vec N}  =  (- f_x, - f_y, 1),
\end{equation}
where $f_x = ({\partial f}/{\partial x})|_{G = 0}$ and $f_y = ({\partial f}/ {\partial y})|_{G =
0}$. Finally, the unit normal vector field to the reflecting surface is given by
\begin{equation}
{\hat N}  =  \frac {( -f_x, -f_y, 1)}{\sqrt{(1 + f_x^2 + f_y^2)}}.
\end{equation}

By using Eqs.\,(1)-(5), a direct computation shows that if ${\vec S} = (s_1, s_2, s_3)$ is the position of the point light source then a light ray that is emitted in the direction ${\hat I}$ and reflected by the arbitrary smooth curved reflector at the point ${\vec r} = (x, y, f(x, y))$ is described by
\begin{eqnarray}
T_1 &=& x+\frac{lh_1}{\alpha},\label{T1} \\
T_2 &=& y+\frac{lh_2}{\alpha},\label{T2}\\
T_3 &=& f(x, y)+ \frac{lh_3}{\alpha}, \label{T3}
\end{eqnarray}
where
\begin{eqnarray}
h_1 &=& (x-s_1)(1-f_x^2+f_y^2)-2f_x[f_y(y-s_2)+s_3-f],\nonumber \\
h_2 &=& (y-s_2)(1+f_x^2-f_y^2)-2f_y[f_x(x-s_1)+s_3-f],\nonumber \\
h_3 &=& (f-s_3)(-1+f_x^2+f_y^2)+2[f_x(x-s_1)+f_y(y-s_2)],\nonumber \\
\alpha &=& (1+f_x^2+f_y^2)\sqrt{(s_1-x)^2+(s_2-y)^2+(s_3-f)^2},
\end{eqnarray}
and $l$, which gives the position of an arbitrary point along the reflected light ray, is such that  $f(x, y) \leq l < \infty$. Furthermore, we assume that $x_{min} \leq x \leq x_{max}$ and $y_{min} \leq y \leq y_{max}$. The values of $x_{min}$, $x_{max}$, $y_{min}$ and $y_{max}$ are determined by the dimensions of the reflecting surface, which we assume  known.

Since we are interested in the intersection of the reflected light rays with an arbitrary plane $z = constant$, then it is convenient to take $T_3 = z_0$, and thus from Eq.\,(\ref{T3}) we have that
\begin{equation}
l = \alpha \left(\frac{T_3 - f(x, y)}{h_3} \right) = \alpha \left(\frac{z_0 - f(x, y)}{h_3} \right),
\end{equation}
so that Eqs.\,(\ref{T1})-(\ref{T3}) can be rewritten in the following form
\begin{eqnarray}
T_1(x, y, z_0) &=& x+[z_0-f(x,y)]\left(\frac{h_1(x,y, s_1, s_2, s_3)}{h_3(x,y, s_1, s_2, s_3)}\right),\nonumber \\
T_2(x, y, z_0) &=& y+[z_0-f(x,y)]\left(\frac{h_2(x,y, s_1, s_2, s_3)}{h_3(x,y, s_1, s_2, s_3)}\right),\nonumber  \\
T_3(x, y, z_0) &=& z_0, \label{T1T2T3}
\end{eqnarray}
where $f(x, y) \leq z_0 < \infty$.

Before continuing it is important to explain the geometrical meaning of the parametric map given by Eqs.\,(\ref{T1T2T3}). For this end, we assume that the position of the point light source is fixed. Then for fixed values of $x$ and $y$; that is, for a point on the reflecting surface, $z = f(x, y)$, Eqs.\,(\ref{T1T2T3}) describe a line segment, which starts at the point ($x$, $y$, $f(x,y)$) and goes to infinity as $z_0$ does. Therefore, as $x$ and $y$ take all their allowed values, this map describes a family of line segments that start at the points of the reflecting surface and go to infinity. This family of line segments are the reflected light rays. On the other hand, from a mathematical point of view, Eqs.\,(\ref{T1T2T3}) describe the parametric form of a map between two subsets of ${\cal R}^3$, where ($x$, $y$, $z_0$) are local coordinates of the domain space and ($T_1$, $T_2$, $T_3$) are local coordinates of the target space. Equivalently, Eqs.\,(\ref{T1T2T3}), can be seen as a \textit{one-parameter} family of maps between subsets of ${\cal R}^2$, where each map is characterized by a particular value of $z_0$. This family is explicitly given by
\begin{eqnarray}
T_1(x, y, z_0) &=& x+[z_0-f(x,y)]\left(\frac{h_1(x,y, s_1, s_2, s_3)}{h_3(x,y, s_1, s_2, s_3)}\right),\nonumber \\
T_2(x, y, z_0) &=& y+[z_0-f(x,y)]\left(\frac{h_2(x,y, s_1, s_2, s_3)}{h_3(x,y, s_1, s_2, s_3)}\right). \label{T1T2}
\end{eqnarray}
Each member of the family, characterized by a specific value of $z_0$, maps points of the reflecting surface to points on the plane $z = z_0$. It is important to remark that in Eqs.\,(\ref{T1T2}), $z_0$ is not a coordinate as it is in Eqs.\,(\ref{T1T2T3}), but it is a parameter characterizing a particular member of the family of maps, and is such that $f(x, y) \leq z_0 < \infty$.

The sets of Eqs.\,(\ref{T1T2T3}) and (\ref{T1T2}) are those maps we were referring to in the introduction, which describe the evolution of the reflected light rays by the arbitrary smooth surface after being emitted by the point light source. We remark that because we have used the reflection law only one time to obtain Eqs.\,(\ref{T1T2T3}), and, therefore,  Eqs.\,(\ref{T1T2}), these equations describe the evolution of the light rays that have experienced only one reflection before leaving the smooth surface. In this work we assume that the parameters that characterize the optical system under study are such that this condition is fulfilled by the reflected light rays.

\begin{figure}[h]
\begin{center}
\includegraphics[height=6cm]{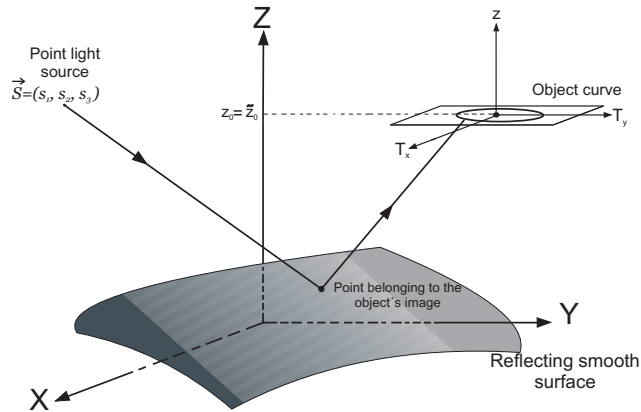}
\caption{\footnotesize Schematic drawing of the two sets of coordinate systems used to compute the image of a one-dimensional object obtained by reflection on an arbitrary smooth surface locally given by $z = f(x, y)$. We also have included an emitted light ray such that its associated reflected light ray connects a point of the smooth surface with a point of the one-dimensional object. The point, on the smooth surface, where the emitted light ray is reflected belongs to the image of the one-dimensional object.}\label{f:5}
\end{center}
\end{figure}

Assuming that in the plane $z = \tilde{z}_0$  a coordinate system ($T_x$, $T_y$) is introduced, with origin at ($0$, $0$, $\tilde{z}_0$) such that
$T_x$ and $T_y$ are parallel to the axes $x$ and $y$ respectively (see figure \ref{f:5}), a one-dimensional object lying on this plane can be described in a parametric way by
\begin{eqnarray}
T_x &=& \Gamma(\sigma), \nonumber\\
T_y &=& \Sigma(\sigma), \label{Objeto}
\end{eqnarray}
where $\sigma$ is a parameter that labels the points on the object. If we eliminate the parameter $\sigma$ we have that the object could be described by
\begin{eqnarray}
T_x &=& \Lambda(T_y)\label{Objeto2}.
\end{eqnarray}
Therefore, the image of this object that an observer, with position  $\vec S = (s_1, s_2, s_3)$, may see on the surface of reflection is given by all the points of the form ($x$, $y$, $f(x,y)$) such that $x$ and $y$ are solutions to
\begin{eqnarray}
T_x(\sigma) &=& x+[\tilde{z}_0-f(x,y)]\left(\frac{h_1(x,y, s_1, s_2, s_3)}{h_3(x,y, s_1, s_2, s_3)}\right),\nonumber \\
T_y(\sigma) &=& y+[\tilde{z}_0-f(x,y)]\left(\frac{h_2(x,y, s_1, s_2, s_3)}{h_3(x,y, s_1, s_2, s_3)}\right).\label{Imagen}
\end{eqnarray}

In other words, a point on the reflecting surface belongs to the image of the object (\ref{Objeto}) if it can be connected, via a reflected light ray, with a point of that object (see figure \ref{f:5}). In the explicit examples that we present later on, we will not plot the image on the reflecting surface, but on the plane $z = 0$. This is because when one takes a photograph of an object its image is normally printed on a plane. Therefore, in our examples, the image will be given by all the points ($x$, $y$, $0$) such that  $x$ and $y$ satisfy Eqs.\,(\ref{Imagen}).

Equations \,(\ref{Imagen}) together with the caustic associated with the reflected light rays will allow us to study the change in the topology of the image of an arbitrary one-dimensional object obtained under reflection on an arbitrary smooth surface.\\

\section{ Computation of the caustic}

Since to study the change of image topology of an arbitrary one-dimensional object obtained by reflection it is required to compute the caustic associated with the reflected light rays described by  Eqs.\,(\ref{T1T2T3}) or equivalently by Eqs.\,(\ref{T1T2}), in this section, following our previous work\, \cite{Gilberto1, Gilberto2}, we review the computation of the caustic by using Eqs.\,(\ref{T1T2T3}) and furthermore we point out its relationship to the caustic associated with Eqs.\,(\ref{T1T2}). To this end, we introduce the following:

\textbf{Definition:} Let $h:\mathcal{M} \rightarrow \mathcal{N}$ be a differentiable map, with $\mathcal{M}$ and $\mathcal{N}$ differentiable manifolds. The set of points in $\mathcal{M}$ where $h$ is not locally one to one are referred to as its critical set, and the image of the critical set is referred to as the caustic set of $h$ \cite{ArnoldI, ArnoldII,ArnoldIII}. If $\mathcal{M}$ and $\mathcal{N}$ are differentiable submanifolds of ${\cal R}^n$ with local coordinates $(x_i)$ and $(y_j)$ respectively, then locally $h$ is given by
\begin{equation}
y_i = h_i(x_j), \hspace{1cm} {\rm where} \hspace{1cm} i,j= 1,…,n. \label{yi}
\end{equation}
Therefore, in this case the critical set is obtained from the condition
\begin{equation}
J\equiv \det\left(\frac{\partial y_i}{\partial x_j}\right) = 0. \label{Jn}
\end{equation}
In general this condition can be written in the following way
\begin{equation}
F(x_1,..., x_n) = 0, \label{JnF}
\end{equation}
if this equation can be solved, for example, for $x_n$, then locally the critical set of $h$ is given by
\begin{equation}
x_n = g(x_1,..., x_{n-1}), \label{JnF}
\end{equation}
which, in the general case, is a set of surfaces of dimension $n-1$ in the domain space which we are assuming  has dimension $n$. Therefore, the caustic set, which is the image of the critical set, is  obtained substituting Eq.\,(\ref{JnF}) into Eqs.\,(\ref{yi}). That is, the caustic set associated with the map (\ref{yi}) is locally given by
\begin{eqnarray}
y_1 & = &  h_1(x_1,..., x_{n-1}, g(x_1,..., x_{n-1})), \nonumber \\
y_2 & = &  h_2(x_1,..., x_{n-1}, g(x_1,..., x_{n-1})), \nonumber \\
    & \cdot & \nonumber \\
    & \cdot & \nonumber \\
    & \cdot & \nonumber \\
y_n & =  & h_n(x_1,..., x_{n-1}, g(x_1,..., x_{n-1})).
\end{eqnarray}

In accordance with the above definition, the critical set of the map given by Eqs.\,(\ref{T1T2T3}), that is, the set of points in the domain space with coordinates ($x$, $y$, $z_0$) such that the map is not locally one to one, is obtained from the following condition
\begin{equation}
J (x, y, z_0) = \det\left(\frac{\partial(T_1, T_2, T_3)}{\partial(x, y, z_0)} \right) =\left(\frac{\partial T_1}{\partial x} \right) \left(\frac{\partial T_2}{\partial y}\right) - \left(\frac{\partial T_1}{\partial y} \right) \left(\frac{\partial T_2}{\partial x}\right) = 0. \label{Ja}
\end{equation}
By using Eqs.\,(\ref{T1T2T3}) a direct computation shows that this condition is equivalent to
\begin{equation}
J(x, y, z_0) = H_2(x, y)\left(\frac{z_0- f}{h_3}\right)^2 + H_1(x, y)\left(\frac{z_0-f}{h_3}\right)+H_0(x, y)=0, \label{Ja3}
\end{equation}
where
\begin{eqnarray}
H_2(x, y) &=& \vec h\cdot\left[\left( \frac{\partial \vec h}{\partial x} \right)\times \left( \frac{\partial \vec h}{\partial y} \right)\right],\nonumber \\
H_1(x, y) &=&  \vec h\cdot \left[\left( \frac{\partial \vec r}{\partial x} \right)\times \left( \frac{\partial \vec h}{\partial y} \right)+\left( \frac{\partial \vec h}{\partial x} \right)\times \left( \frac{\partial \vec r}{\partial y} \right) \right],  \nonumber \\
H_0(x, y) &=& \vec h\cdot\left[\left( \frac{\partial \vec r}{\partial x} \right)\times \left( \frac{\partial \vec r}{\partial y} \right)\right],
\end{eqnarray}
with
\begin{eqnarray}
\vec r &=&(x, y, f(x, y)), \nonumber \\
\vec h &=&(h_1, h_2, h_3).
\end{eqnarray}
From Eq.\,(\ref{Ja3}) we find that the critical set associated with the map given by Eqs.\,(\ref{T1T2T3}) is given by
\begin{equation}
z_0 = z_{0 \pm}(x, y) \equiv f + h_3 \left(\frac{-H_1\pm\sqrt{H_1^2-4H_2H_0}}{2H_2}\right). \label{Cc3}
\end{equation}
Therefore, the caustic set associated with the map given by Eqs.\,(\ref{T1T2T3}), which by definition is obtained by substituting Eq.\,(\ref{Cc3}) into    Eqs.\,(\ref{T1T2T3}), can be written in the following form
\begin{equation}
{\vec T}_{c\pm}=\vec r + \left(\frac{-H_1\pm\sqrt{H_1^2-4H_2H_0}}{2H_2}\right)\vec h. \label{Cau3}
\end{equation}
It is important to remark that this equation is equivalent to that obtained by Shealy and Burkhard\,\cite{Shealy1, Shealy2} by using a different procedure. From this last equation it is clear that the caustic set or simply the caustic associated with the reflected light rays described by Eqs.\,(\ref{T1T2T3}), in general, is composed by two branches, which for very particular forms of the reflecting surface and particular positions of the point light source reduce to a single point. For example, this happens when the reflecting surface is  part of a perfect spherical mirror and the point light source is located at its center of curvature. In the general case the two branches of the caustic are two-dimensional surfaces which when they are stable under small deformations of the reflecting surface and the position of the point light source, they locally have singularities of well known types: the swallowtail, the pyramid or elliptic umbilic and the purse or hyperbolic umbilic \cite{ArnoldI, ArnoldII,ArnoldIII}. The fact that the caustic might not be stable under small deformations of the reflecting surface and the position of the point light source is related to the symmetries of the system formed by the reflecting surface and the point light source.\\

Since we are assuming that $x_{min} \leq x \leq x_{max}$ and $y_{min} \leq y \leq y_{max}$, then in our case the caustics will be located within a finite region of the space. In particular there exist real numbers  $z_{cmin\pm }$ and $z_{cmax\pm }$ such that $z_{cmin\pm } \leq T_{3c\pm} \leq z_{cmax\pm }$.

Now we compute the critical and caustic sets associated with the one-parameter family of maps given by Eqs.\,(\ref{T1T2}) and we point out their relationship to those associated with the map (\ref{T1T2T3}). To this end, we select the map given by Eqs.\,(\ref{T1T2}) when $z_0 = \tilde{z}_0 = constant$. That is,
\begin{eqnarray}
T_1(x, y) &=& x+[\tilde{z}_0-f(x,y)]\left(\frac{h_1(x,y, s_1, s_2, s_3)}{h_3(x,y, s_1, s_2, s_3)}\right),\nonumber \\
T_2(x, y) &=& y+[\tilde{z}_0-f(x,y)]\left(\frac{h_2(x,y, s_1, s_2, s_3)}{h_3(x,y, s_1, s_2, s_3)}\right).\label{T1T2z0}
\end{eqnarray}
where $\tilde{z}_0$ is a real number such that $f(x, y) \leq \tilde{z}_0 \leq \infty$. Remember that this map sends points from the reflecting surface to points on the plane $z_0 = \tilde{z}_0 = constant$. The critical set associated with this map is obtained from the following condition
\begin{equation}
\tilde{J} (x, y) = \det\left(\frac{\partial(T_1, T_2)}{\partial(x, y)} \right) =\left(\frac{\partial T_1}{\partial x} \right) \left(\frac{\partial T_2}{\partial y}\right) - \left(\frac{\partial T_1}{\partial y} \right) \left(\frac{\partial T_2}{\partial x}\right) = 0, \label{Ja}
\end{equation}
or equivalently from
\begin{equation}
\tilde{J} (x, y) = H_2(x, y)\left(\frac{\tilde{z}_0- f}{h_3}\right)^2 + H_1(x, y)\left(\frac{\tilde{z}_0-f}{h_3}\right)+H_0(x, y)=0. \label{Ja2}
\end{equation}
In this equation $\tilde{z}_0$ is a number, it is not a variable as $z_0$ is in Eq.\,(\ref{Ja3}). Observe that the relationship between $\tilde{J} (x, y)$ and  $J(x, y, z_0)$  is given by $\tilde{J} (x, y) = J(x, y, \tilde{z}_0)$.  This means that only when
\begin{equation}
z_{cmin\pm } \leq \tilde{z}_0 \leq z_{cmax\pm},
\end{equation}
the map given by Eqs.\,(\ref{T1T2z0}) will be not locally one to one.  Therefore, if $\tilde{z}_0$ satisfies the above condition, from Eq.\,(\ref{Ja2}) we have that locally the critical set associated with the map (\ref{T1T2z0}) may be written in the  form
\begin{equation}
y = \Psi(x, \tilde{z}_0). \label{Jacc}
\end{equation}
In this last equation we have included $\tilde{z}_0$, which is a number, only to show that it is the critical set associated with the map given by Eqs.\,(\ref{T1T2z0}). The caustic set associated with this particular map, which is obtained by substituting Eq.\,(\ref{Jacc}) into Eq.\,(\ref{T1T2z0}), is given by
\begin{eqnarray}
T_{1c}(x) &=& x+[\tilde{z}_0-f(x,\Psi)]\left(\frac{h_1(x,\Psi, s_1, s_2, s_3)}{h_3(x,\Psi, s_1, s_2, s_3)}\right),\nonumber \\
T_{2c}(x) &=& \Psi+[\tilde{z}_0-f(x,\Psi)]\left(\frac{h_2(x,y, s_1, s_2, s_3)}{h_3(x,\Psi, s_1, s_2, s_3)}\right),\label{T1T2cau}
\end{eqnarray}
in the general case, it is a curve or family of curves in the plane $z_0 = \tilde{z}_0 = constant$. For very particular cases it reduces to a point. When the caustic is stable under small deformations of the optical system formed by the reflecting surface and the point light source, then in accordance with the Whitney's singularity theory\,\cite{ArnoldI, ArnoldII,ArnoldIII}, it locally has singularities of fold or cusp type.

From the computations presented in this section it is clear that the caustic associated with the map (\ref{T1T2z0}) given by Eq.\,(\ref{T1T2cau}) is equal to the intersection of the caustic associated with the map (\ref{T1T2T3}), given by Eq.\,(\ref{Cau3}), with the plane $z_0 = \tilde{z}_0 = constant$.

Before closing this section, we explain the geometrical meaning of the caustic associated with the evolution of the reflected light rays by the smooth arbitrary surface when the point source is located at any position of the space. To this end, consider the pencil of light rays reflected by the differential surface $dxdy$ of the reflector. From Eqs.\,(\ref{T1T2z0}) we have that when $\tilde{z}_0 = f$ the cross-sectional area of this pencil of rays is exactly $dxdy$, and as the light rays evolve, this area is given by $dT_1dT_2 = \mid \tilde{J}(x, y) \mid dxdy = \mid \tilde{J}(x, y, \tilde{z}_0)\mid dxdy$. If ($x$, $y$) belongs to the critical set of the map given by Eqs.\,(\ref{T1T2z0}) then $\tilde{J}(x, y) = J(x, y, \tilde{z}_0) = 0$ and therefore in that case the cross-sectional area of the pencil of rays collapses to zero. This result shows that the caustic is defined by the focusing region associated with the reflected light rays and therefore also can be defined as the singularities of the flux density\,\cite{Shealy1, Shealy2}. Finally, we remark that in the literature one can find explicit expressions and plots of the caustic associated with particular reflectors and certain positions of the point light source\,\cite{Shealy3, Theocaris, CorderoIII}.

\section{ Equations to study the change of image topology}

In this section we summarize the two sets of equations that allow us to study, in an exact way, the changes in the topology of the image, of an arbitrary object, obtained by reflections on an arbitrary smooth surface:\\

\noindent $\clubsuit$ The image of an arbitrary one-dimensional object, locally given by Eqs.\,(\ref{Objeto}) and located on the plane $z = \tilde{z}_0 = constant$, obtained by reflection on the smooth reflecting surface locally given by $z = f(x, y)$, is computed by solving the Eqs.\,(\ref{Imagen}); that is
\begin{eqnarray}
T_x(\sigma) &=& x+[\tilde{z}_0-f(x,y)]\left(\frac{h_1(x,y, s_1, s_2, s_3)}{h_3(x,y, s_1, s_2, s_3)}\right),\nonumber \\
T_y(\sigma) &=& y+[\tilde{z}_0-f(x,y)]\left(\frac{h_2(x,y, s_1, s_2, s_3)}{h_3(x,y, s_1, s_2, s_3)}\right),\label{ImagenF}
\end{eqnarray}
for $x$ and $y$. The solution must be such that $x_{min} \leq x \leq x_{max}$ and $y_{min} \leq y \leq y_{max}$.\\

\noindent $\clubsuit \clubsuit$ The caustic associated with the reflected light rays described by the map (\ref{T1T2z0}) is given by the intersection of the caustic (\ref{Cau3}), that is
\begin{equation}
{\vec T}_{c\pm}=\vec r + \left(\frac{-H_1\pm\sqrt{H_1^2-4H_2H_0}}{2H_2}\right)\vec h, \label{Cau3F}
\end{equation}
and the plane $z = \tilde{z}_0 = constant$. Or equivalently given by Eq.\,(\ref{T1T2cau}).\\

Remember that the map given by Eqs.\,(\ref{T1T2z0}) maps points on the reflecting surface to points on the plane $z = \tilde{z}_0 = constant$. If the plane $z = \tilde{z}_0 = constant$, which contains the object, is outside  the caustic region; that is, there is not intersection between this plane and the caustic given by Eq.\,(\ref{Cau3F}), then the map (\ref{T1T2z0}) is locally one to one. This means that for each point ($T_x(\sigma)$, $T_y(\sigma)$, $\tilde{z}_0$) on the object there is a unique solution ($x$, $y$) to Eqs.\,(\ref{ImagenF}) and therefore, under this condition, the observer will register a single image. In other words, the object and its associated image, in this case, have the same topology. Now we assume that the plane $z = \tilde{z}_0 = constant$ is located at the caustic region given by Eq.\,(\ref{Cau3F}). In this second case, the map given by Eqs.\,(\ref{T1T2z0}) is not locally one to one at those points that belong to the intersection of the caustic and the plane $z = \tilde{z}_0 = constant$. Therefore, if the object lying on this plane is outside  the caustic then the observer will register a single image, but if the object reaches the caustic in such a way that they become tangent to each other; that is, there is a touch between them, then the object and its associated image do not have the same topology and therefore the observer can see multiple images corresponding to a single object (see the examples in the next section).\\

We close this section with the following observations:\\

\noindent $\spadesuit$) The set of Eqs.\,(\ref{ImagenF}) have been reported in the literature by Cordero, Cornejo and Cardona\,\cite{CorderoI} in the context of the Ronchi and Hartmann tests. These authors showed that Eqs.\,(\ref{ImagenF}) allow to describe the main features of both tests. In particular, they found that the ronchigram associated with an arbitrary reflecting surface, when the point source is located at an arbitrary position and the Ronchi ruling is at the plane $z = \tilde{z}_0 = constant$ with its rulings parallel to the $x$ axis, is given by the level curves of $T_y(\sigma)$.

\noindent $\spadesuit\spadesuit$) The caustic, Eq.\,(\ref{Cau3F}), has also been reported in \cite{Gilberto2, Shealy1, Shealy2} and, in particular, it was used to compute the circle of least confusion associated with a rotationally symmetric mirror when the point light source is located on the optical axis\,\cite{Gilberto1, Gilberto2, CorderoIII, Hosken}.

\noindent $\spadesuit\spadesuit \spadesuit $)  Though the two sets of equations (\ref{ImagenF}) and  (\ref{Cau3F}) have been reported, to our knowledge, they have not been used to explain the structure of the ronchrigrams when the Ronchi ruling is located at the caustic region. The main contribution of this work is to realize that the procedure implemented by Berry is equivalent to that used in the Ronchi test and therefore that the closed loops fringes observed in the Ronchi pattern when the Ronchi ruling is located at the caustic region can be explained by using the caustic touching theorem. (The present work can be considered as a logical continuation of research reported in Refs.\, \cite{CorderoI, Gilberto1, Gilberto2}.)\\

\section{Examples: Spherical mirror with the point light source on the optical axis}

In order to illustrate everything we have presented in the previous sections we assume that the reflecting smooth surface is a part of a spherical mirror with radius $r$ and diameter $D$ given by
\begin{equation}
z = f(x, y) =  r - \sqrt{r^2 - x^2 - y^2},
\end{equation}
where $- D/2 \le x \leq D/2$ and  $- D/2 \le y \leq D/2$.  Equivalently, this equation can be written in the following form
\begin{equation}
z = f(\rho) =  r - \sqrt{r^2 - \rho^2},
\end{equation}
where $\rho =\sqrt{x^2 + y^2}$ is such that $0 \leq \rho \leq \rho_{max} = D/2$. For this case
\begin{eqnarray}
T_1(x, y, z_0) & = & x [1 + (z_0 - r + \sqrt{r^2 - \rho^2})G(\rho)] , \nonumber \\
T_2(x, y, z_0) & = & y [1 + (z_0 - r + \sqrt{r^2 - \rho^2}) G(\rho)], \nonumber \\
T_2(x, y, z_0) & = & z_0, \label{T1T2T3esf}
\end{eqnarray}
where
\begin{eqnarray}
G(\rho) =  \frac{2 (r -
s)(r^2  - \rho^2) - r^2 \sqrt{r^2 - \rho^2}}
{\sqrt{r^2 - \rho^2}[2 \rho^2 (r - s) + r^2(s - r + \sqrt{r^2 - \rho^2})]}.
\end{eqnarray}
A direct computation shows that the two branches of the critical set associated with the map Eq.\,(\ref{T1T2T3esf}) are given by
\begin{eqnarray}
z_{0-} & = & \frac{r[2 r^3 + 2 s \rho^2 - 2 r^2 (s + \sqrt{r^2 - \rho^2}) - r (2 \rho^2 - s\sqrt{r^2 - \rho^2})]}{2r^3 - 2r \rho^2
+ 2 s \rho^2 - r^2 (2 s + \sqrt{r^2 - \rho^2})},\nonumber \\
z_{0+} & = & \frac{r^3 (4 r^2 - 5 r s + 2 s^2) - (r - s) [r^2 (4 r - s) + 2 (r - s)\rho^2] \sqrt{r^2 - \rho^2}}
{r^2[3 r^2 - 4 r s + 2 s^2 + 3(s - r)\sqrt{r^2 - \rho^2}]},\nonumber \\ \label{z0masmenos}
\end{eqnarray}
and the corresponding branches of the caustic set by
\begin{eqnarray}
T_{1c-}(x, y) & = & 0,  \nonumber \\
T_{2c-}(x, y) &=& 0,\nonumber  \\
T_{3c-}(x, y) &=& z_{0-}, \label{caumenosE}
\end{eqnarray}
and
\begin{eqnarray}
T_{1c+}(x, y) & = & \frac{2x (r - s)^2  \rho^2}{r^2[3 r^2 - 4 r s + 2 s^2 + 3 (s - r) \sqrt{r^2 - \rho^2}]}, \nonumber \\
T_{1c+}(x, y) & = & \frac{2y (r - s)^2  \rho^2}{r^2[3 r^2 - 4 r s + 2 s^2 + 3 (s - r) \sqrt{r^2 - \rho^2}]}, \nonumber \\
T_{3c+}(x, y) &=& z_{0+}. \label{caumasE}
\end{eqnarray}

\begin{figure}[h]
\begin{center}
\includegraphics[height=8cm]{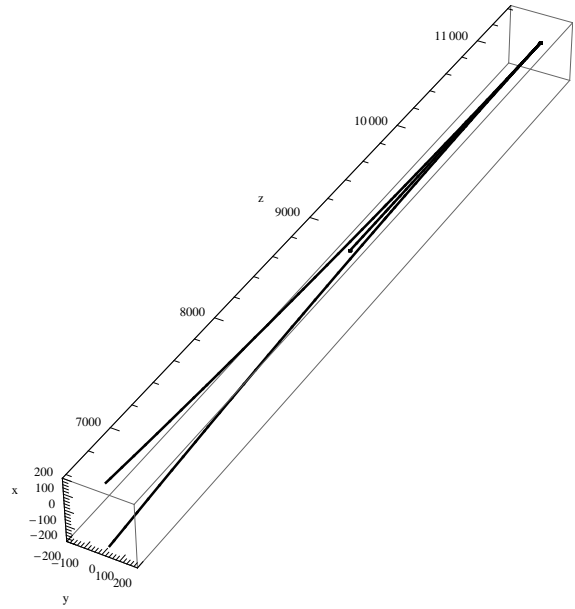}
\caption{\footnotesize
Intersection of the caustic given by Eqs.\,(\ref{caumenosE}) and (\ref{caumasE}) with the plane $y = 0$, for the special case $r = 2415$ mm, $D = 1470$ mm $ = 2 \rho_{max}$, $z = z_0 = 11000$ mm and the point light source is located at ($0$, $0$, $1350$ mm).}\label{caumas}
\end{center}
\end{figure}

\begin{figure}[h]
\begin{center}
\includegraphics[height=5cm]{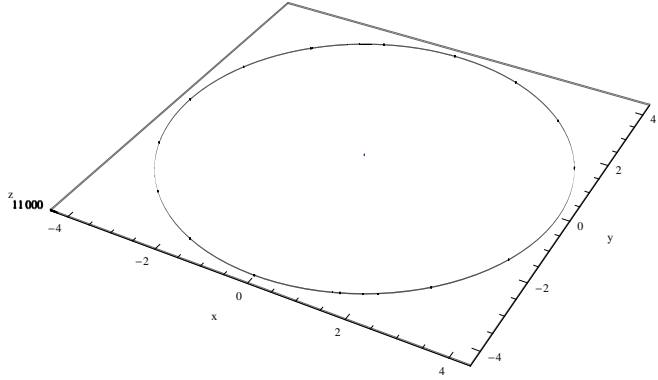}
\caption{\footnotesize Intersection of the two branches of the caustic with the plane $z = 11000$.}\label{cauint}
\end{center}
\end{figure}

Equations\,(\ref{caumenosE}) describe a line segment on the optical axis, that is on the $z$ axis. This part of the caustic corresponds to the fact that the rays that leave the points in a circle on the mirror with its center on the optical axis converge at a point. All the convergent points of all the rays that leave all the possible circles on the mirror with centers on the optical axis form this part of the caustic (the line segment). By contrast Eqs.\,(\ref{caumasE}), in general, describe a surface of revolution with a degenerated singularity of cusp type. If we intersect this surface with a plane containing the $z$ axis then we obtain a curve with a singularity of cusp type. However, if we intersect it with a plane perpendicular to the $z$ axis then we obtain a circle of radius $\sqrt{T_{1c+}^2 + T_{2c+}^2}$. It is an example where one of the branches of the caustic is not a two-dimensional surface, this fact is related to the axial symmetry of the optical system under study. Even more, as it is clear from Eqs.\,(\ref{z0masmenos})-(\ref{caumasE}) for the very particular case $s = r$, that is, when the point light source is located at the center of the spherical mirror, the two branches of the caustic reduce to an isolated point.

By using Eqs.\,(\ref{z0masmenos}), a direct computation shows that when $s \geq r$ then
\begin{eqnarray}
T_{3c-}(\rho_{max}) \leq & T_{3c-}(\rho) & \leq \frac{rs}{2s -r}, \nonumber \\
T_{3c+}(\rho_{max}) \leq & T_{3c+}(\rho) & \leq \frac{rs}{2s -r}.
\end{eqnarray}
Therefore, there is an interval on the $z$ axis where $T_{3c-}(\rho) = T_{3c+}(\rho)$ (see figure\,\ref{caumas}). This means that, in this case, the intersection of the caustic with a plane $z = z_0 = constant$ such that $z_0 = T_{3c-}(\rho) = T_{3c+}(\rho)$, is a circle and an isolated point (see figure\,\ref{cauint}). The circle is obtained from the intersection between the plane and that part of the caustic that is a surface of revolution, while the point is obtained from the intersection between the plane and the line segment corresponding to the other branch of the caustic.

For this case, the one-parameter family of maps between points on the spherical mirror and points of an arbitrary plane $z = z_0 = constant$, is explicitly given by:
\begin{eqnarray}
T_1(x, y) & = & x [1 + (z_0 - r + \sqrt{r^2 - \rho^2})G(\rho)] , \nonumber \\
T_2(x, y) & = & y [1 + (z_0 - r + \sqrt{r^2 - \rho^2}) G(\rho)].\label{T1T2E}
\end{eqnarray}
From the previous discussion we have, if $z_0$ is such that $z_0 = T_{3c-}(\rho) = T_{3c+}(\rho)$ then the caustic associated with this map is a circle and a point which coincides with the center of the circle.

If in the plane $z = z_0 = constant$, we have a one-dimensional object given by (\ref{Objeto}), then its image is obtained by solving, for $x$ and $y$, the following set of equations:
\begin{eqnarray}
T_x(\sigma) & = & x [1 + (z_0 - r + \sqrt{r^2 - \rho^2})G(\rho)] , \nonumber \\
T_y(\sigma) & = & y [1 + (z_0 - r + \sqrt{r^2 - \rho^2}) G(\rho)]. \label{Imagenc}
\end{eqnarray}
Given $T_x(\sigma)$, $T_y(\sigma)$, $z_0$ and $r$, the image on the spherical mirror is given by all the points ($x$, $y$, $r - \sqrt{r^2 - x^2 - y^2}$), where $x$ and $y$ are solutions to Eqs.\,(\ref{Imagenc}). In this work, instead of plotting the image on the smooth reflecting surface, we will plot it on the $z = 0$ plane.

\begin{figure}[h]
\begin{center}
\includegraphics[width=12cm]{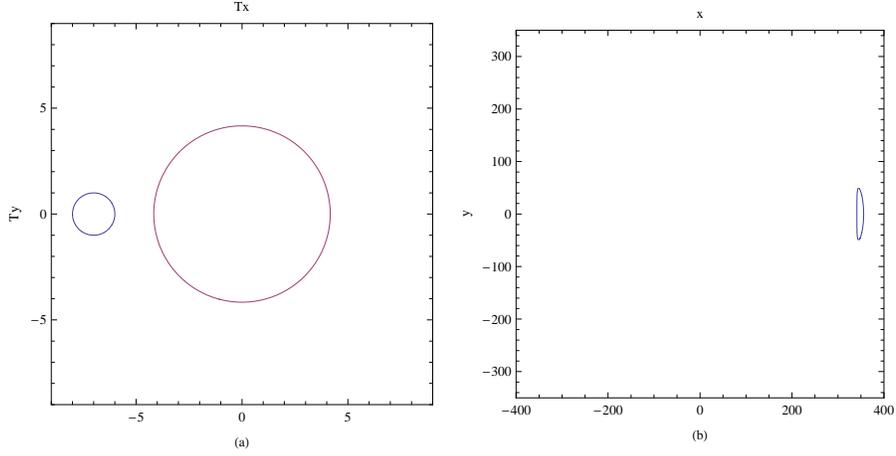}
\caption{\footnotesize (a) Object space and (b) Image space for $k = -7$.}\label{ObImI}
\end{center}
\end{figure}

\begin{figure}[h]
\begin{center}
\includegraphics[width=12cm]{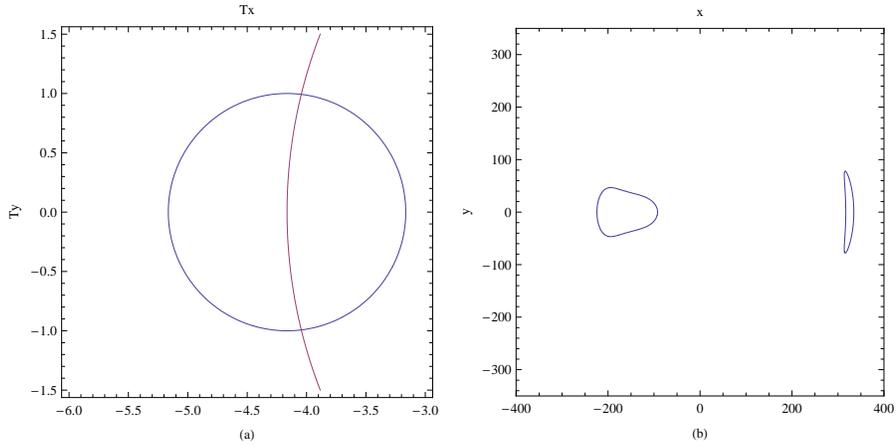}
\caption{\footnotesize (a) Object space and (b) Image space for $k = -R_c$.}\label{ObImIII}
\end{center}
\end{figure}

For the explicit computations we take: $r = 2415$ mm, $D = 1470$ mm $ = 2 \rho_{max}$, the point light source at ($0$, $0$, $1350$ mm) and $z = z_0 = 11000$ mm as the plane where the object is located at. Using these data, in  figure\,(\ref{caumas}), we present the intersection  of the two branches of the caustic given by Eqs.\,(\ref{caumenosE}) and (\ref{caumasE}) with the plane $y = 0$. While in figure\,(\ref{cauint}) we present the intersection of these two branches of the caustic with the plane $z = 11000$, that is, with the plane where the one-dimensional object is located at. The intersection is a circle of radius $R_c = 4.1643$ mm and its center. (In the plots presented in the next section corresponding to the object space, the plane $z = 11000$, we only show that part of the caustic that corresponds to the circle, we are not including the part corresponding to the isolated point.)

\subsection{First example: Circular Object}

In this example we assume that the object is a circle of radius $1$  given by
\begin{equation}
(T_x - k)^2 + T_y^2 = 1,\label{circuloOb}
\end{equation}
where $k$ is a real constant. That is, we are interested in obtaining the image of the object that is a circle with radius equal to $1$ with center at  ($k$, $0$), which is located in the plane  $z = 11000$. From Eqs.\,(\ref{Imagenc}) and (\ref{circuloOb}) we obtain that for a fixed value of $k$, the image of the circle (\ref{circuloOb}) in the plane $z = 0$ is given by all the values of $x$ and $y$ such that:
\begin{eqnarray}
&&\{x [1 + (11000 - r + \sqrt{r^2 - \rho^2})G(\rho)] - k\}^2 \nonumber \\
&& \hspace{3cm}  +   y^2 [1 + (11000 - r + \sqrt{r^2 - \rho^2}) G(\rho)]^2 = 1. \label{circulo}
\end{eqnarray}

\begin{figure}[h]
\begin{center}
\includegraphics[width=12cm]{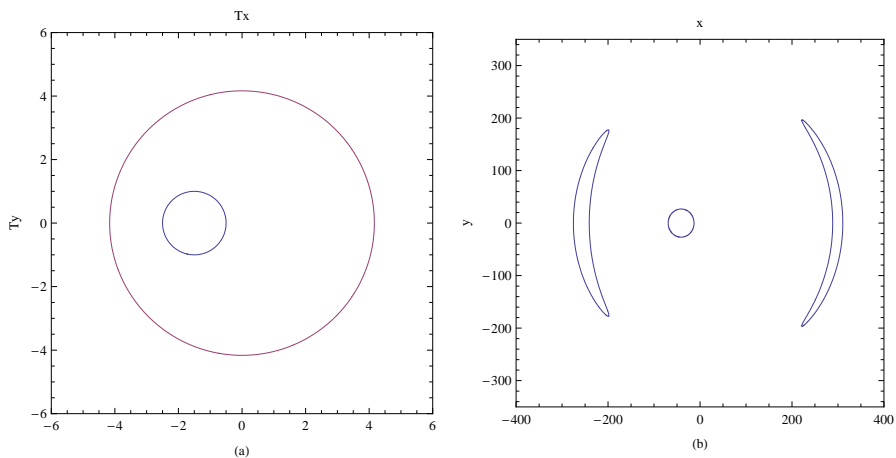}
\caption{\footnotesize (a) Object space and (b) Image space for $k = -1.5$.}\label{ObImIX}
\end{center}
\end{figure}

We have written a computer program in \textit{Mathematica} to study the change of topology of the image for this case and those cases presented later on. That is, we give a particular value to $k$, and in the object space ($T_x$, $T_y$, 11000) we plot the object (circle of radius $1$) and the caustic (remember that we only will plot the circle of radius $R_c$). After that, the set of Eqs.\,(\ref{circulo}) are solved for $x$ and $y$ under the conditions $-\rho_{max} = -735 \leq x \leq \rho_{max} = 735$ and $-\rho_{max} = -735 \leq y \leq \rho_{max} = 735$. Finally, these values obtained for $x$ and $y$ are plotted in the plane $z= 0$. In what follows, we present some plots where the change of topology is clearly observed.

\begin{figure}[h]
\begin{center}
\includegraphics[width=12cm]{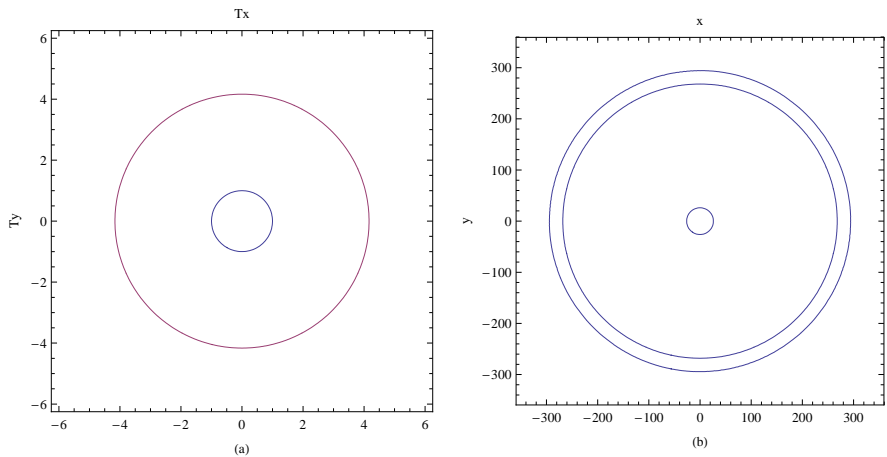}
\caption{\footnotesize (a) Object space and  (b) Image space for $k = 0$.}\label{ObImXI}
\end{center}
\end{figure}

In  figure \,(\ref{ObImI}) we show the object and image for $k = -7$. In this case, we only get one image because the map that sends points from the spherical mirror to points of the object is one to one; that is, the object is out of the caustic region. In accordance with the caustic touching theorem we will observe a change of image topology when the object and the caustic touch  each other, that is when they become tangent to each other. If we assign values to $k$ from $-7$  to $7$, then the first change of topology is obtained when $k = -R_c -1$ (remember that $R_c = 4.1643$). For this value of $k$ a new image appears from an isolated point. In figure\,(\ref{ObImIII}) we present the case corresponding to $k = -R_c$, here it is clearly observed the second image. There is another change of topology when  $k = -R_c + 1$, for this value of $k$ a third image appears. In figure \,(\ref{ObImIX})  we present the case $k = -1.5$, where the third image is clearly observed. Other change of topology is obtained for $k = -1$ because the caustic, in the plane $z = 11 000$, is constituted by the circle of radius $R_c$ and its center. For this particular value of $k$ the object is tangent to that part of the caustic that corresponds to the isolated point. For $k = 0$, the images are three concentric circles (see figure {\ref{ObImXI}). Now it is clear that there will be other changes of image topology when $k = 1, R_c -1$ and  $R_c + 1$.

\begin{figure}[h]
\begin{center}
\includegraphics[width=12cm]{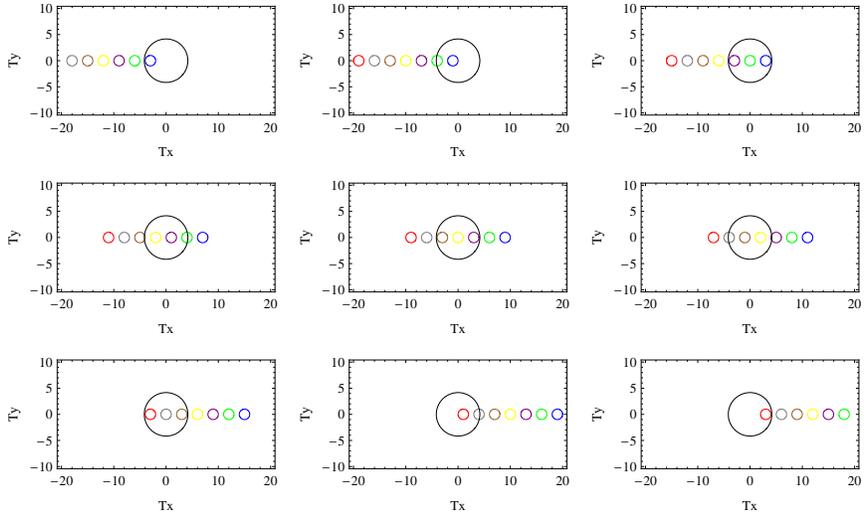}
\caption{\footnotesize  Set of circles with their centers on $T_x$.}\label{OLC}
\end{center}
\end{figure}

Finally, in figure (\ref{OLC}) we show a set of circles with their centers on the $T_x$ axis in the object space and in figure (\ref{ILC}) we present the corresponding images. While in figure (\ref{OCC}) we show a family of objects corresponding to concentric circles with different positions with respect to the caustic, and in figure (\ref{ICC}) we present their associated images. It is important to remark that Murty and Shoemarker\,\cite{CCG} have presented the theory of a method to test optical systems, which is similar to that of Ronchi, but instead of using straight lines for the grating they used concentric circles. They have presented characteristic patterns for the usual aberrations of optical systems. In some of these patterns it is possible to see the change of image topology. However this fact was not pointed out by these authors.

\begin{figure}[h]
\begin{center}
\includegraphics[width=12cm]{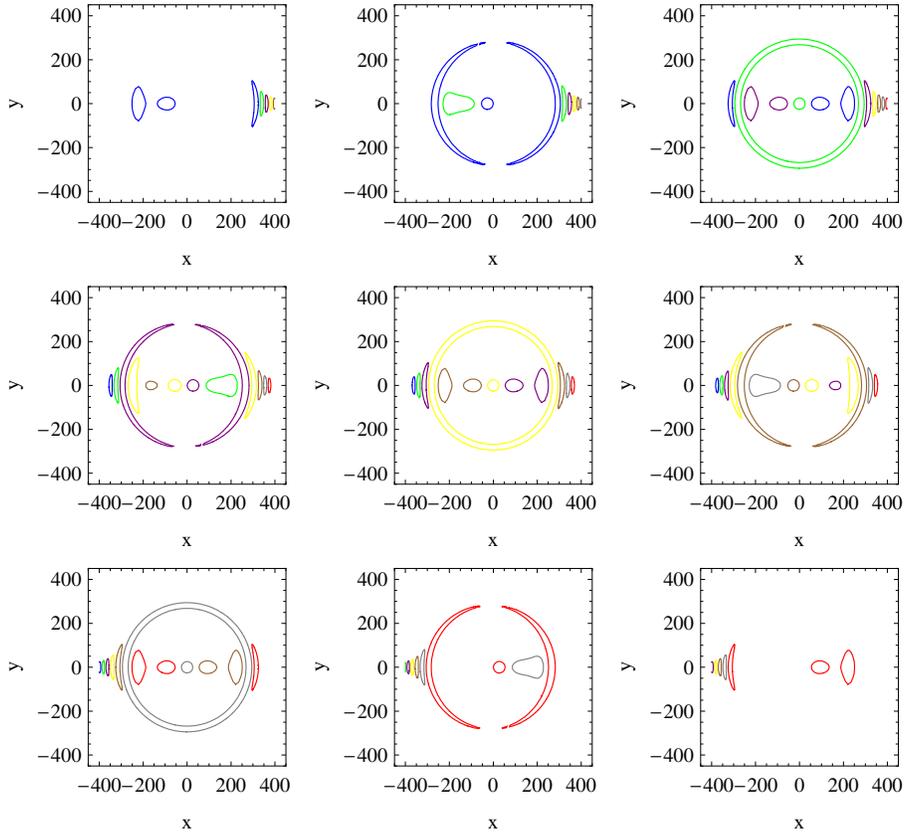}
\caption{\footnotesize Image of the set of  circles with their centers on $T_x$.}\label{ILC}
\end{center}
\end{figure}

\begin{figure}[h]
\begin{center}
\includegraphics[width=12cm]{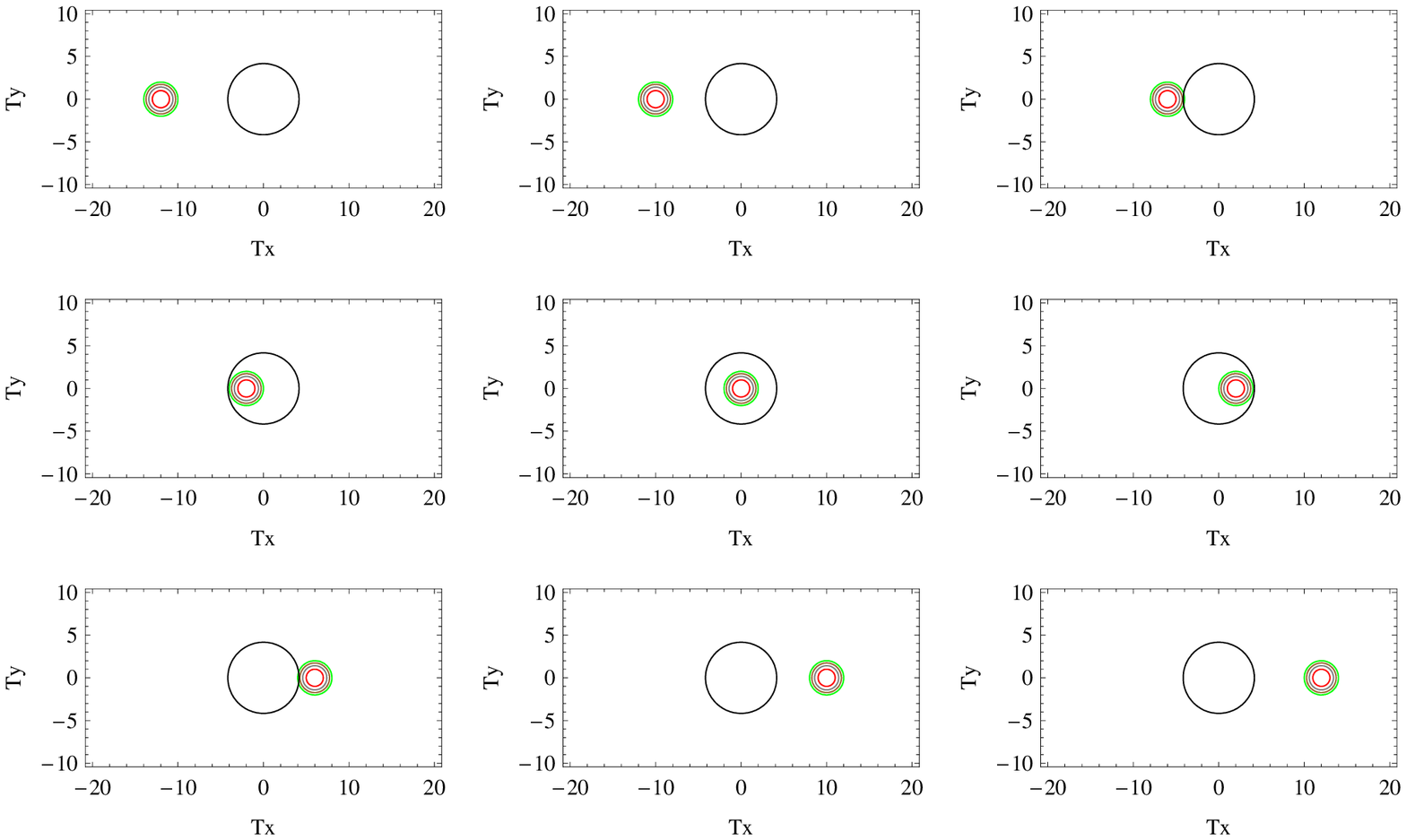}
\caption{\footnotesize Set of concentric circles with their centers on $T_x$.}\label{OCC}
\end{center}
\end{figure}

\begin{figure}[h]
\begin{center}
\includegraphics[width=12cm]{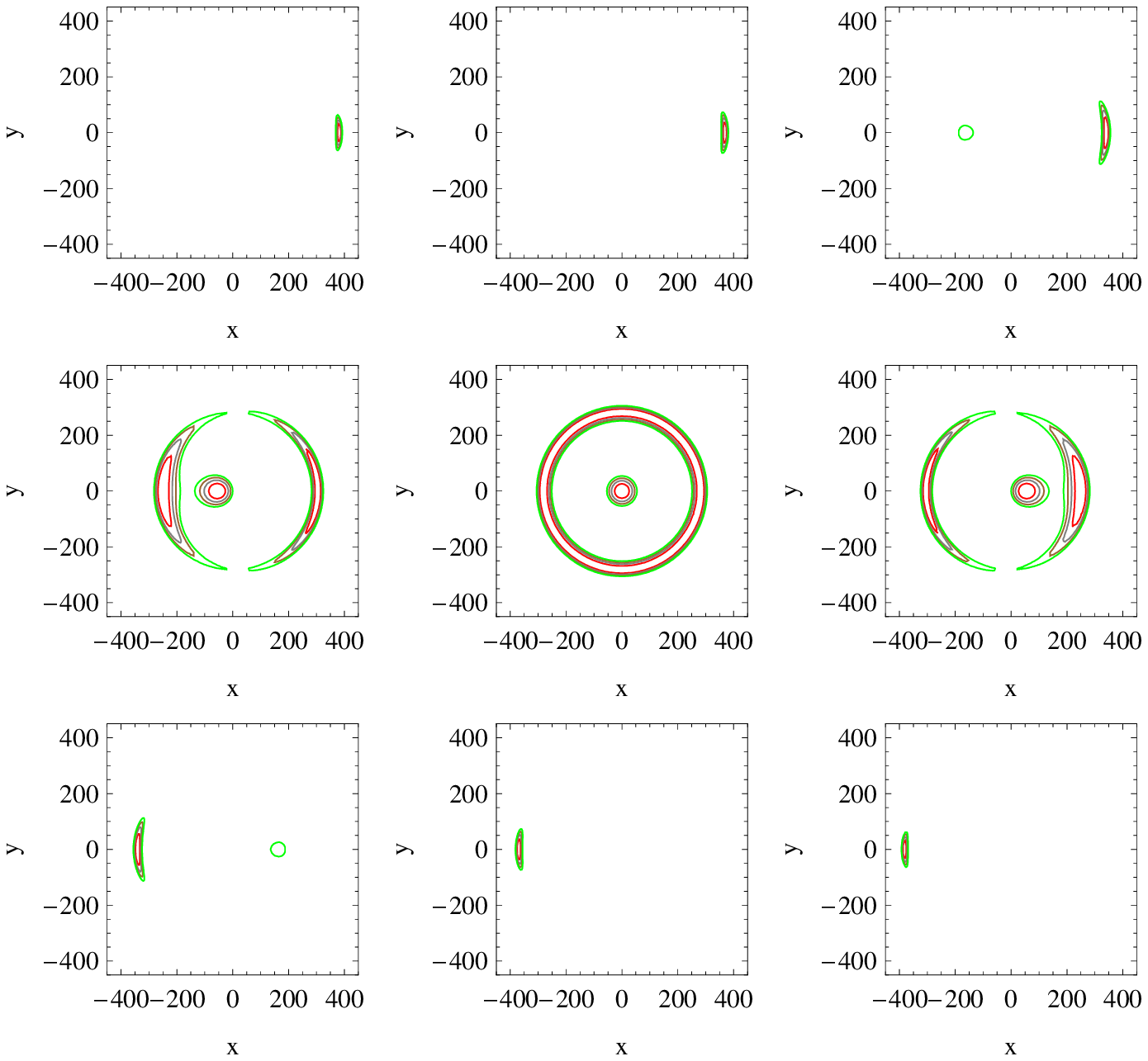}
\caption{\footnotesize Image of the set of concentric circles with their centers on $T_x$.}\label{ICC}
\end{center}
\end{figure}

\clearpage

\subsection{Second example: Linear object}

\begin{figure}[h]
\begin{center}
\includegraphics[width=12cm]{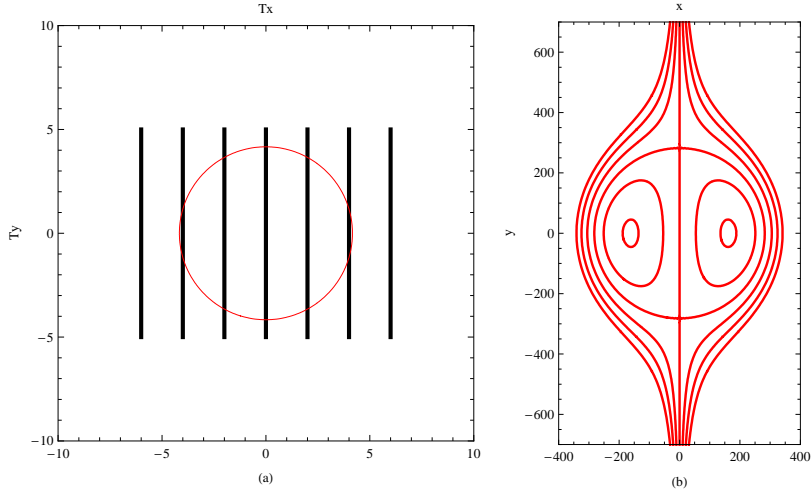}
\caption{\footnotesize (a) Object space: set of line segments parallel to the $T_y$ axis with $T_y \in [-6, 6]$ and the caustic, which is a circle of radius $R_c = 4.1643$mm and its center. (b) Image space: the corresponding images. In the Ronchi test the set of lines (a) is the grating or Ronchi ruling and its image (b) is referred to as the associated Ronchigram.}\label{REJRON}
\end{center}
\end{figure}

In this second example, which we believe is the most relevant due to its connection to the famous Ronchi test used to test in particular conic mirrors, we assume that the object is a line segment parallel to the $T_y$ axis. Actually in the Ronchi test the grating is formed by ruling bands, for our purposes we can assume that the line segments that we consider, coincide with the center of the ruling bands of the Ronchi ruling. That is, we assume that the object is given by
\begin{equation}
T_x = k,\label{lineaOb}
\end{equation}
where $k$ is a real constant. In other words, we are interested in obtaining the level curves associated with the function $T_1$, given in Eqs.\,(\ref{T1T2E}). From Eqs.\,(\ref{Imagenc})  and (\ref{lineaOb}) we have that for $k$ fixed, the image in the plane $z = 0$ is given by all the points of the form ($x$, $y$, $0$) such that $x$ and $y$ satisfy the following equation
\begin{eqnarray}
x [1 + (11000 - r + \sqrt{r^2 - \rho^2})G(\rho)] = k. \label{ImagenL}
\end{eqnarray}

\begin{figure}[h]
\begin{center}
\includegraphics[width=12cm]{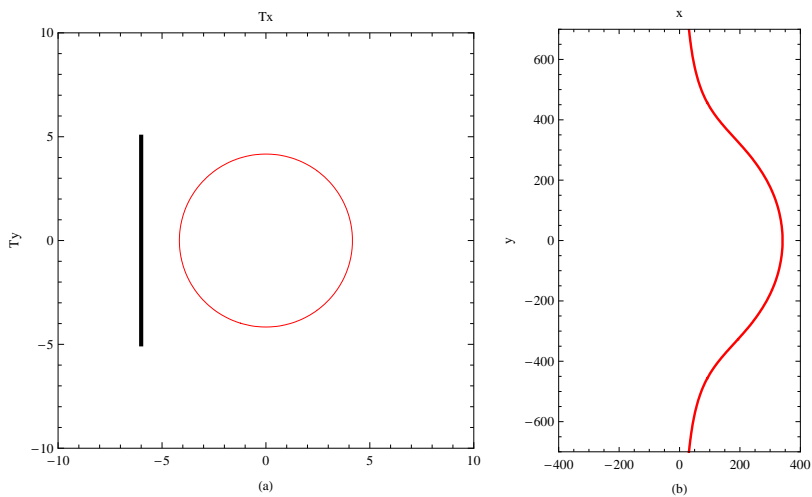}
\caption{\footnotesize (a) Object space and (b) Image space for $k = - 6$.}\label{ObImk-6}
\end{center}
\end{figure}

In exactly the same way as in the previous case, to study the change in the topology of the image, we give different values to $k$ in the interval [$-6$,  $6$]. In figure\,\ref{REJRON}(a) we show the caustic and the object for different values of $k$. In  figure\,\ref{REJRON}(b) can be observed the corresponding images. In the Ronchi test  figure\,\ref{REJRON}(a) corresponds to the Ronchi ruling and  figure\,\ref{REJRON}(b) corresponds to what is referred to as the ideal ronchigram associated with a spherical mirror. The ronchigram, figure\,\ref{REJRON}(b), has been reported in the literature, but its closed loops fringes never had been explained by using the caustic touching theorem as we are doing in this work.

\begin{figure}[h]
\begin{center}
\includegraphics[width=10cm]{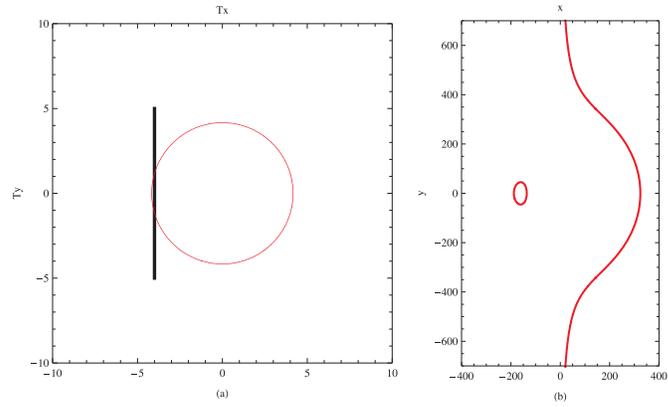}
\caption{\footnotesize (a) Object space and (b) Image space for $k = - 4$.}\label{ObImk-4}
\end{center}
\end{figure}

We start with the analysis of the change of the image topology: in figure\,\ref{ObImk-6}(a) we show the object (line segment given by Eq.\,(\ref{ImagenL}) with $k = -6$) and the caustic, which are in the plane $z = z_0 = 11000$. In figure\,\ref{ObImk-6}(b) we show its image, which is not a segment of line but it is a distortion of it. For this value of $k$, that is for $k = -6$, we obtain only one image because any point of the line segment can be reached by only one reflected light ray. The first change in the image topology is obtained when the line segment is tangent to that part of the caustic corresponding to the circle of radius $R_c$, that is when $k = -R_c$, at this value of $k$  a new image is born as an isolated point. The second image that has born is visible for $k = -4$, as can be seen in figures\,\ref{ObImk-4}(a) and \ref{ObImk-4}(b). In figure\,(\ref{ObImk-2}) we have presented the case $k = - 2$. The following change of topology occurs when $k = 0$, it is because the center of the circle with radius $R_c$ is also part of the caustic. In figures\,\ref{ObImIV}(a)-\ref{ObImIV}(h) we show the object space and the image space for $k = 0, 2, 4, 6$, respectively. Now it is clear that for this example, if $k \in (-R_c, 0)$ the object and its image have the same topology. The second image that was born when $k = - R_c$  transforms into a circle when $k = 0$. For $k \in (0, R_c)$ we have two images. Finally, the last change in the image topology is obtained when $k = R_c$. For this value of $k$ one of the images reduces to an isolated point, so that for $k > 0$ we only have a single image.

Remember that when $k = -R_c$  in the image space there appears a new image from an isolated point, which as $k$ goes to zero  transforms into a circle, something similar happens for $0< k < R_c$, but in this second case the circle reduces to an isolated point when $k = R_c$. In figure\,(\ref{ImagenRE2}) we have isolated some of these loop images for some values of $k$. It is important to remark that this pattern was obtained by Ronchi\,\cite{Ronchi} when he was studying a lens with a remarkable aberration. He also obtained the pattern corresponding to $k = 0$.

Finally, in the set of figures (\ref{ORR}) and (\ref{IRR}) we show the object space and the image space associated with a set of objects emulating the Ronchi ruling and their associated ronchigrams.

\begin{figure}[h]
\begin{center}
\includegraphics[width=10cm]{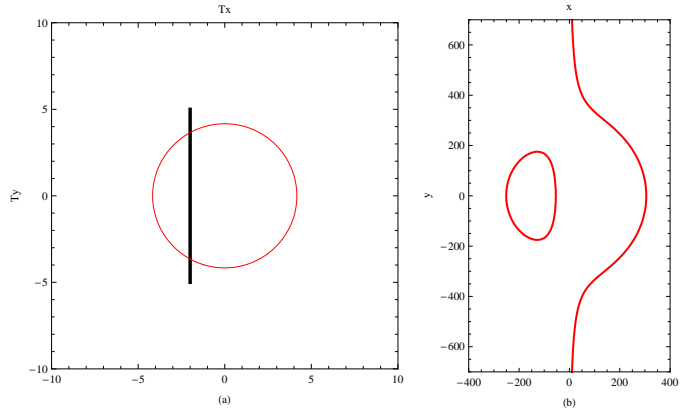}
\caption{\footnotesize (a) Object space and (b) Image space for $k = -2$.}\label{ObImk-2}
\end{center}
\end{figure}

\begin{figure}[h]
\begin{center}
\includegraphics[width=15cm]{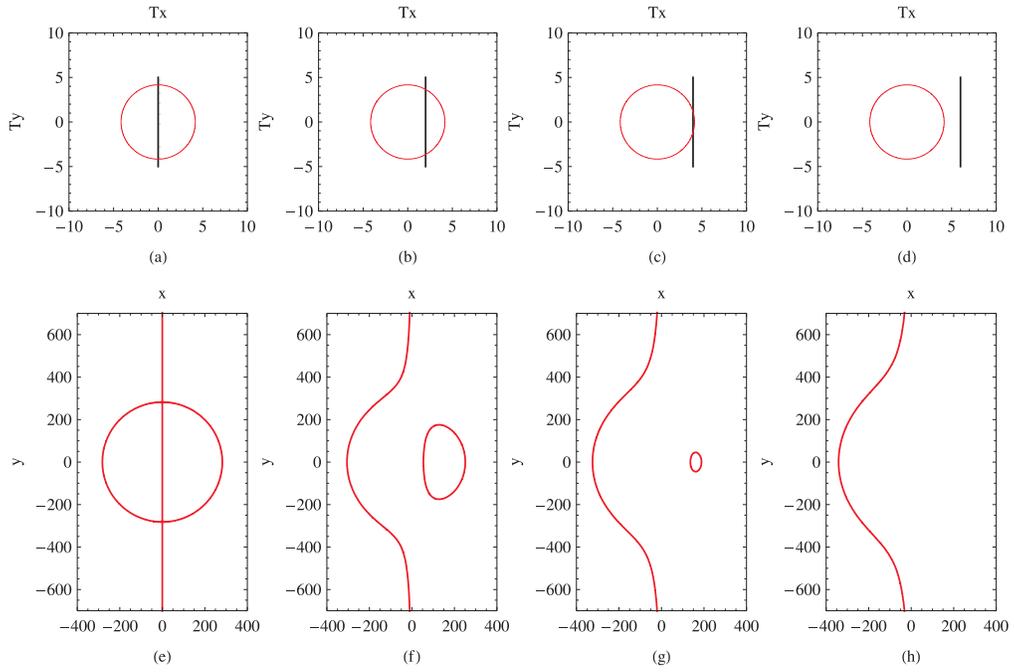}
\caption{\footnotesize (a)-(d) Object space and (e)-(h) Image space for $k = 0, 2, 4, 6$, respectively.}\label{ObImIV}
\end{center}
\end{figure}

\begin{figure}[t]
\begin{center}
\includegraphics[width=8cm]{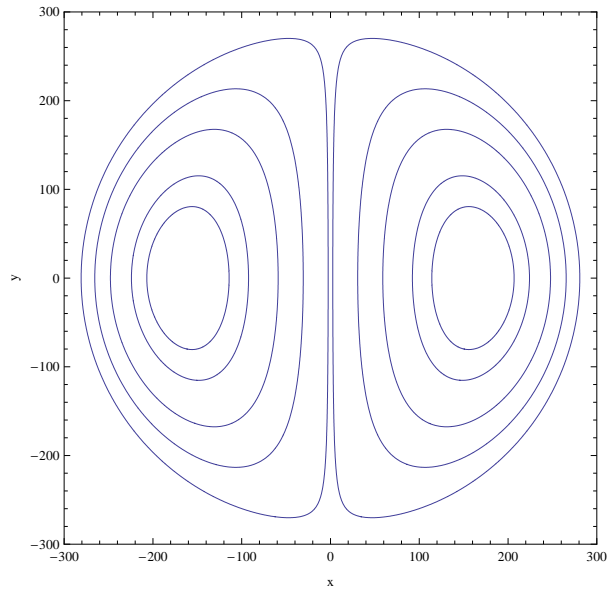}
\caption{\footnotesize Pattern obtained by Ronchi.}\label{ImagenRE2}
\end{center}
\end{figure}

\begin{figure}[t]
\begin{center}
\includegraphics[width=16cm]{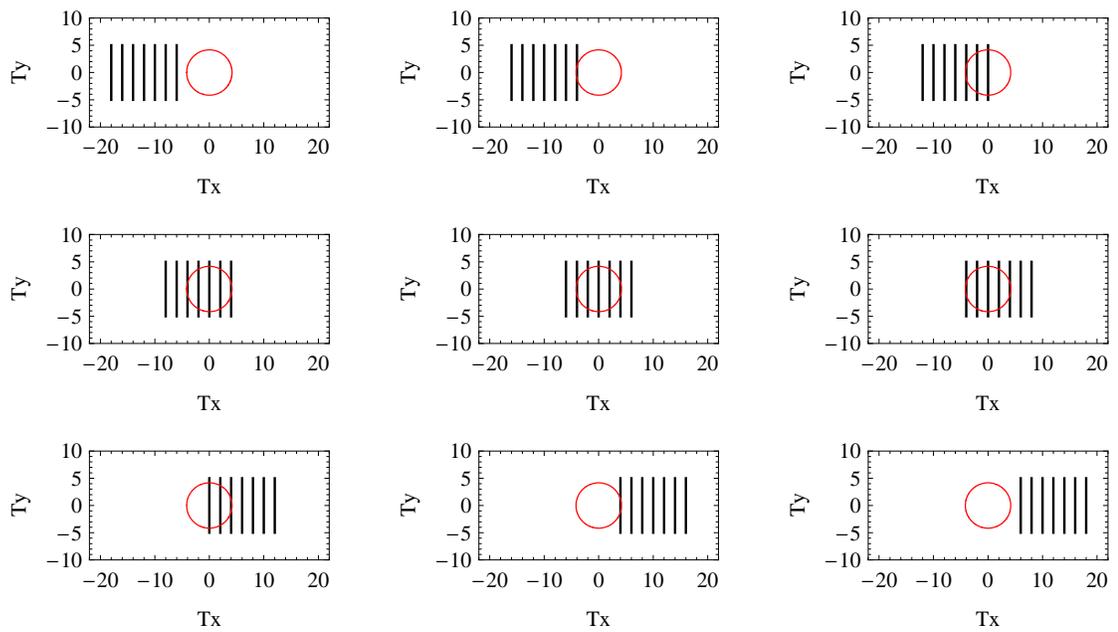}
\caption{\footnotesize Simulation of the Ronchi ruling and the caustic curve.}\label{ORR}
\end{center}
\end{figure}

\begin{figure}[h]
\begin{center}
\includegraphics[width=14cm]{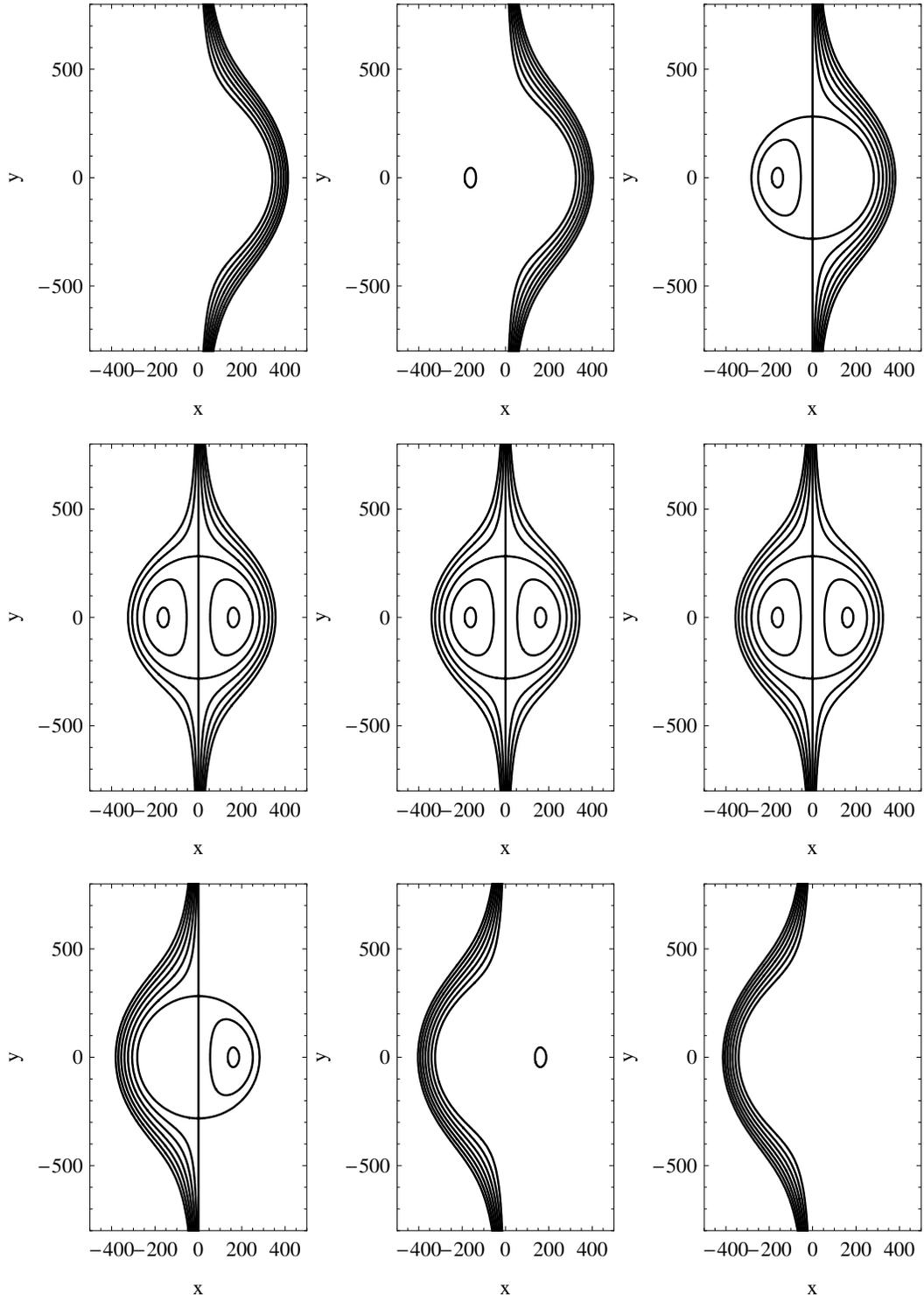}
\caption{\footnotesize Image of the Ronchi rulings or ronchigrams.}\label{IRR}
\end{center}
\end{figure}

\clearpage

\section{Conclusions}

In this work we have obtained an exact set of equations to study the change of image topology of an arbitrary object obtained by reflection on an arbitrary smooth surface. We have pointed out that the equations that allow to compute the image of an arbitrary one-dimensional object have been reported in the literature\,\cite{CorderoI} in the context of the Ronchi and Hartmann tests. However, to our knowledge, nobody has remarked that the closed loops fringes observed in the ronchigram when the Ronchi ruling is located at the caustic place are due a disruption of images or fringes. We believe that the main contribution of this work is to realize that the procedure developed by Berry to study the changes of image topology of the Sun disk, is equivalent to that used in the simulation of ideal ronchigrams. From this observation it is clear that the caustic plays a major role in describing correctly this kind of pattern. We claim that the results established in this work could provide a description of the patterns obtained in other tests that use gratings.

Our general results were illustrated by several examples when the reflecting surface is a spherical mirror and the point light source (in the Ronchi test) or observing eye (in Berry's procedure), is located on the optical axis. We believe, it could be worthwhile to study the change of image topology when the point light source is out of the optical axis and the reflecting surface is a conic reflector.

We remark that analogous results can be obtained for lenses. In a future paper we will report these results.

\section{Acknowledgments} The authors thank M. Berry and an unknown referee for helpful comments on the manuscript. E. Rom\'an-Hern\'andez was supported by a CONACyT scholarship and G. Silva-Ortigoza acknowledges financial support from SNI (M\'exico) and CONACyT.

\end{document}